\documentclass{jps-cp}
\usepackage{txfonts} 
\usepackage{graphicx}
\usepackage{hyperref}
\usepackage{footmisc}

\makeatletter
\def\ext@figure{}
\makeatother

\title{Extragalactic Sources and Propagation of UHECRs}

\author{Arjen \textsc{van Vliet}$^{1}$, Rafael \textsc{Alves Batista}$^{2}$ and G\"unter \textsc{Sigl}$^{3}$}

\inst{$^{1}$Department of Astrophysics/IMAPP, Radboud University, P.O. Box 9010, 6500 GL Nijmegen, The Netherlands \\
$^{2}$Department of Physics - Astrophysics, University of Oxford, DWB, Keble Road, OX1 3RH, Oxford, UK \\
$^{3}$II. Institut f\"ur Theoretische Physik, University of Hamburg, Luruper Chaussee 149, 22761 Hamburg, Germany}

\email{a.vanvliet@astro.ru.nl}

\recdate{April 17, 2017}

\abst{
With the publicly available astrophysical simulation framework for propagating extraterrestrial UHE particles, CRPropa 3, it is now possible to study realistic UHECR source scenarios including deflections in Galactic and extragalactic magnetic fields in an efficient way. Here we discuss three recent studies that have already been done in that direction. The first one investigates what can be expected in the case of maximum allowed intergalactic magnetic fields. Here is shown that, even if voids contain strong magnetic fields, deflections of protons with energies $\gtrsim 60 \; \text{EeV}$ from nearby sources might be small enough to allow for UHECR astronomy. The second study looks into several scenarios with a smaller magnetization focusing on large-scale anisotropies. Here is shown that the local source distribution can have a more significant effect on the large-scale anisotropy than the EGMF model. A significant dipole component could, for instance, be explained by a dominant source within 5~Mpc distance. The third study looks into whether UHECRs can come from local radio galaxies. If this is the case it is difficult to reproduce the observed low level of anisotropy. Therefore is concluded that the magnetic field strength in voids in the EGMF model used here is too low and/or there are additional sources of UHECRs that were not taken into account in these simulations.
}

\kword{UHECR propagation, cosmic-ray sources, extragalactic magnetic field models}

\begin{document}
\maketitle

\section{Introduction}

Cosmic rays with energy $E \gtrsim 5$~EeV (the ankle) are expected to come mostly from extragalactic sources. However, which extragalactic sources actually produce these ultra-high energy cosmic rays (UHECRs) is unclear. Due to deflections in the extragalactic magnetic field (EGMF) and in the Galactic magnetic field (GMF) the cosmic rays do not point back straight to their sources. For higher and higher energies, however, the magnitude of these deflections decreases. Additionally, for $E \gtrsim 40$~EeV, interactions of UHECRs with the cosmic microwave background (CMB) limit the distance to the sources from which these cosmic rays can originate to $\sim 100$~Mpc. There are not many sources within this distance which might be able to accelerate particles up to such high energies. Therefore, a high level of anisotropy in the cosmic-ray sky at the highest energies could be expected.

The strange thing is that most measurements on UHECRs are compatible with a completely isotropic sky. One exception at large scales might be a dipole moment above 8~EeV for which an indication was found by the Pierre Auger Observatory (Auger)~\cite{Aab:2016ban, ThePierreAuger:2014nja} or above 10~EeV as detected by Auger and Telescope Array (TA) combined~\cite{Deligny:2015vol, Aab:2014ila}. No other deviations from isotropy at large scales are observed by Auger or TA for moments beyond the dipole moment or for lower energies. At intermediate scales TA has reported the observation of a `hot spot' of events with an energy above 57 EeV at a $3.4\sigma$ significance level~\cite{Kawata:2015whq, Abbasi:2014lda}. Searches for small to intermediate scale anisotropies by Auger all yielded no significant indication of anisotropy (see e.g. Ref.~\cite{Aublin:2015txt}).

Such a high level of isotropy could possibly be explained by having a rigidity-dependent maximum acceleration energy at the sources, making the mass composition at the highest energies relatively heavy and the maximum energy at the sources relatively low (see e.g. the best-fit results from Ref.~\cite{Aab:2016zth}). A heavy mass composition will increase the deflections by magnetic fields at the highest energies, increasing the level of isotropy. Furthermore, a relatively low maximum rigidity relaxes the constraints on the source energy and allows for a higher number of possible sources within the local universe, which also increases the level of isotropy. Such a heavy mass composition agrees well with recent mass composition measurements by Auger~\cite{Aab:2014kda}. However, especially these mass measurements suffer from relatively low statistics and large uncertainties for $E \gtrsim 40$~EeV. The planned upgrade of the Pierre Auger Observatory~\cite{Aab:2016vlz} will be able to provide significant improvements in that respect. 

To test this hypothesis detailed simulations of the propagation of UHECRs are required including all relevant interactions, realistic models of the extragalactic and Galactic magnetic field and models for the distribution of possible sources. For that purpose we have developed the public\footnote{available from \url{https://crpropa.desy.de}} astrophysical simulation framework for propagating extraterrestrial UHE particles called CRPropa 3~\cite{Batista:2016yrx}. The effects of uncertainties in these kind of simulations of the propagation of UHECRs on predictions of observable quantities is studied extensively in Ref.~\cite{Batista:2015mea}.

\section{CRPropa 3}

\begin{figure}[tbh]
	\begin{centering}
		\includegraphics[scale=0.5]{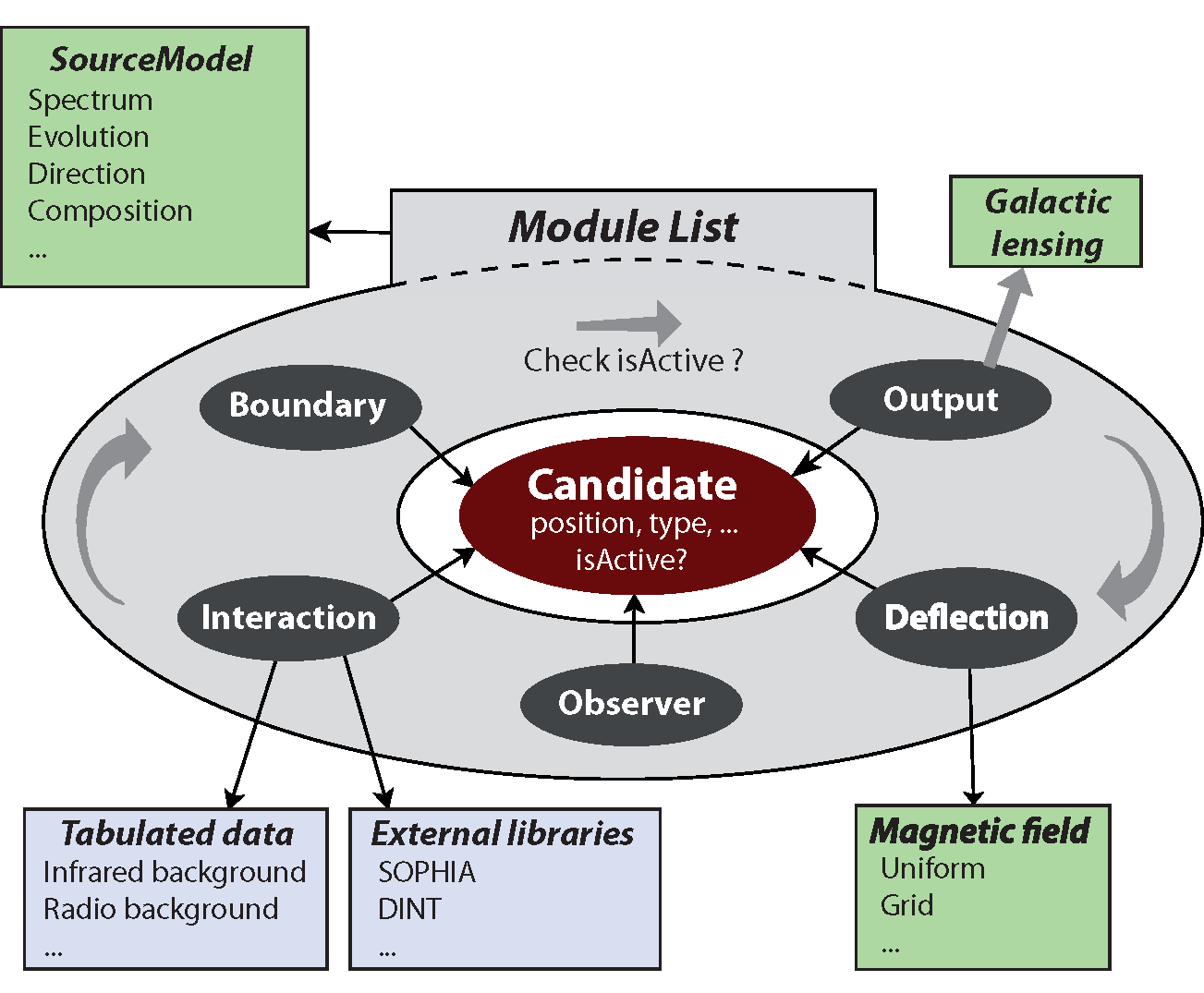}
		\caption{Schematic overview of the modular structure of CRPropa 3, obtained from Ref.~\cite{Batista:2016yrx}. Cosmic-ray candidates are created by a source-model class. The modules only interact with this central cosmic-ray candidate, not with other modules. This continues until a certain break condition is satisfied.}
		\label{fig:visu}		
	\end{centering}
\end{figure}

Version 3 of CRPropa has been officially released recently accompanied by Ref.~\cite{Batista:2016yrx}. It is designed to provide predictions for astrophysical observables of UHE particles in an efficient way. It has a completely modular structure. Each module is independent of other modules allowing for any combination of modules to be selected. This provides multiple use cases, extensive flexibility in testing, changing, replacing or adding specific modules, and the possibility to study individual propagation aspects. The central cosmic-ray candidate class, with which all modules can interact, stores all relevant information about the particle state. Its properties are updated on each step of the propagation until certain observation or boundary conditions are met. Cosmic-ray candidates can be created individually by the user or by a source-model class. A schematic overview of this code structure, obtained from Ref.~\cite{Batista:2016yrx}, is given in Fig.~\ref{fig:visu}.

The dominant energy-loss processes for UHECRs are pair production, photopion production and photodisintegration with the cosmic microwave background (CMB) and the extragalactic UV/optical/IR background light (EBL). Additionally, during these interactions, unstable nuclei can be created so nuclear decay has to be taken into account as well. Modules are available for all these different processes in CRPropa 3. In addition secondary neutrinos as well as photons, electrons and positrons (causing electromagnetic cascades) are created in these interactions. CRPropa also allows for the simulation of the production and propagation of these cosmogenic neutrinos and electromagnetic cascades.

CRPropa 3 can simulate the propagation through both extragalactic magnetic fields (EGMF) and Galactic magnetic fields (GMF). These magnetic fields can be implemented as analytical fields, grid-like fields or fields with complex structures of any scale. An EGMF obtained by constrained large-scale structure (LSS) MHD simulations by Dolag {\it et al.}~\cite{Dolag:2004kp}, including instruction on how to implement it in CRPropa, is available from a link on the CRPropa website. Available models for the GMF within CRPropa are a toroidal halo field model based on Refs.~\cite{Prouza:2003yf, Sun:2007mx}, field models of axisymmetric (ASS) and bisymmetric (BSS) logarithmic spiral shape, the Pshirkov 2011 GMF model~\cite{Pshirkov:2011um} and the JF 2012 GMF model~\cite{Jansson:2012pc, Jansson:2012rt}.

To treat the Galactic propagation efficiently, after the extragalactic propagation is done, a so-called lensing technique has been developed. By backtracking particles through a specific GMF model until the edge of the Galaxy a `Galactic lens' is created. This Galactic lens is a matrix which transforms incoming extragalactic arrival directions to arrival directions observed at Earth. This method assumes that no energy-loss processes take place during the Galactic propagation. A safe assumption due to the comparatively small travel distances inside our Galaxy. Lenses for the JF 2012 and the BSS model are available from a link on the CRPropa website.

In Ref.~\cite{Batista:2016yrx} example simulations of cosmic-ray arrival directions including deflections in extragalactic and Galactic magnetic fields are discussed. In those scenarios the sources are distributed randomly following the LSS simulations of Ref.~\cite{Dolag:2004kp} with a source density of $10^{-3}$~Mpc$^{-3}$. The structure of the EGMF is obtained from Ref.~\cite{Dolag:2004kp} as well. The magnetic field strength, however, is derived from the relation between matter density and magnetic field strength obtained from simulations by Ref.~\cite{Sigl:2004yk}. They found on average a stronger magnetic field strength than was obtained in Ref.~\cite{Dolag:2004kp}. For the Galactic propagation the lensing technique was used with the JF 2012 GMF model. This simulation setup resulted in dipole amplitudes in the UHECR arrival direction skymaps of $\sim 6-7\%$ for both injecting protons and iron nuclei at the sources, and for both before and after deflection in the GMF.

\section{Maximum Allowed Intergalactic Magnetic Field}

Large discrepancies exist across cosmological simulations of the magnetized cosmic web, partly due to the different assumptions made (e.g. initial conditions, feedback processes, sub-grid models, etc) and partly due to differences in the numerical treatment (e.g. grid resolution). These discrepancies in the modelling of EGMFs, are a major source of uncertainty when searching for the sources of UHECRs. Early works by Sigl {\it et al.}~\cite{Sigl:2004yk} concluded that deflections in EGMFs are large, whereas Dolag {\it et al.}~\cite{Dolag:2004kp} have argued that deflections are small, both of them making use of cosmological simulations.

\begin{figure}[tbh]
	\includegraphics[width=0.5\columnwidth]{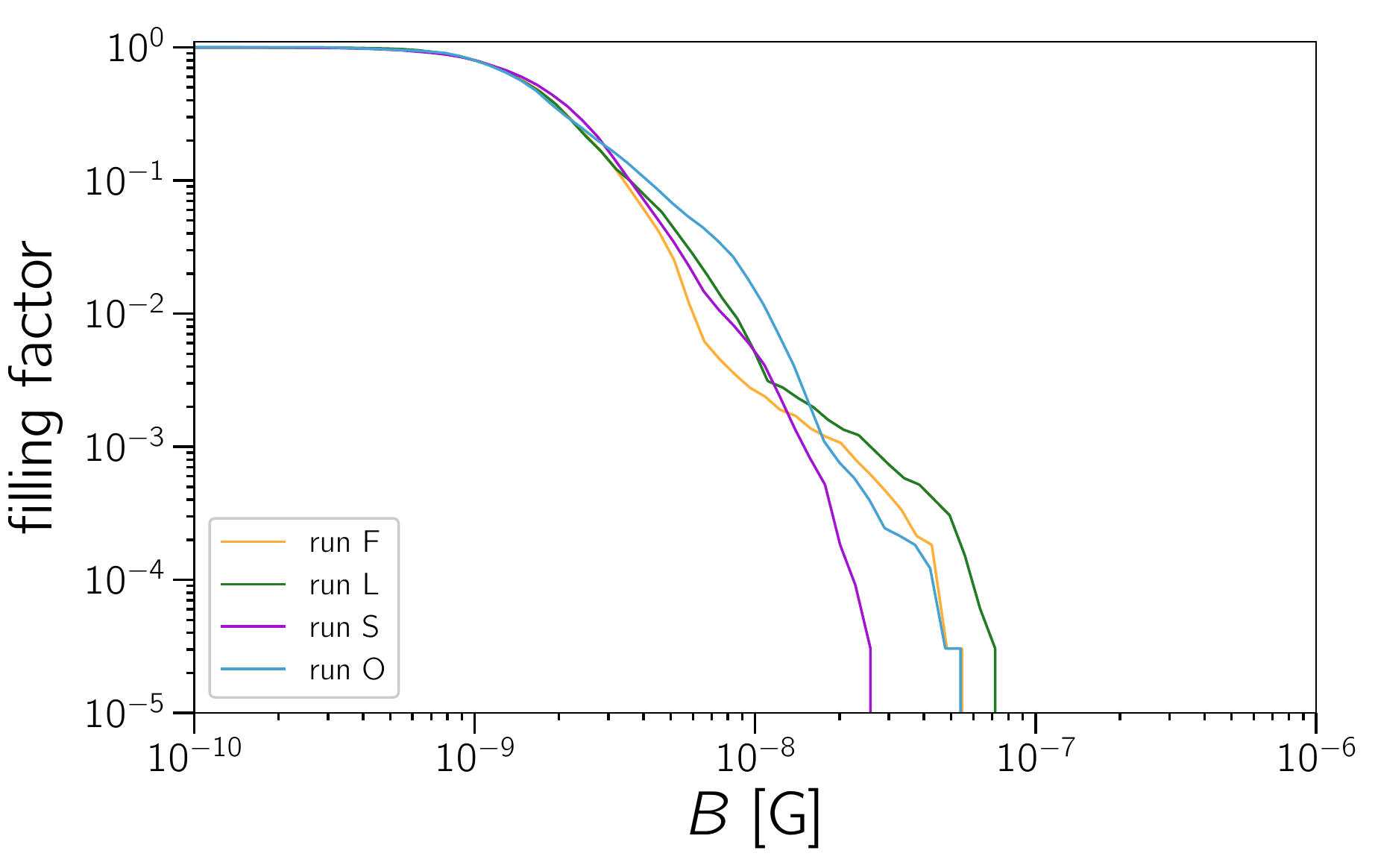}
	\includegraphics[width=0.5\columnwidth]{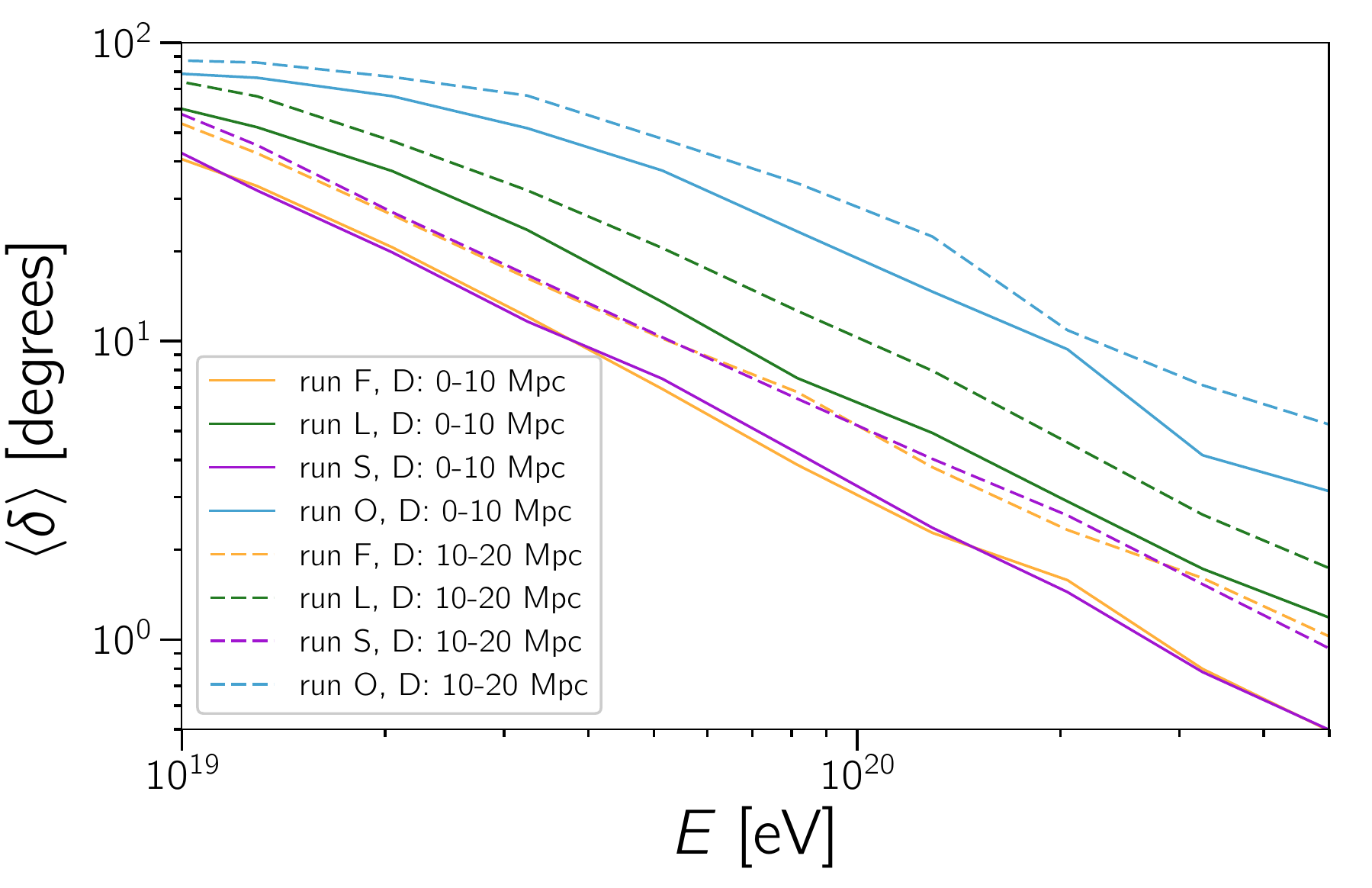}
	\caption{The cumulative filling factors for the magnetic fields of the cosmological simulations are shown in the left panel. The average deflection of protons in the different runs are shown in the right panel for two distance bins: 0-10 Mpc (solid) and 10-20 Mpc (dashed lines). Figures adapted from Ref.~\cite{alvesbatista2017}.}
	\label{fig:maxB}
\end{figure}

Purely magnetohydrodynamical simulations of the cosmic web by Shin {\it et al.}~\cite{shin2017} were used by Alves Batista {\it et al.}~\cite{alvesbatista2017} to study the propagation of UHECRs in the extreme case in which the magnetic field in the voids are close to the upper limit of $\sim 1 \; \text{nG}$. These simulations have a comoving volume of $(200 h^{-1} \; \text{Mpc})^3$. Four configurations for the power spectrum of the seed magnetic field were used, namely run F (fiducial), run L (less power over small scales), run S (less power over large scales), and run O (power only on large scales). They are exactly the same, apart from the initial magnetic power spectrum. The cumulative filling factors for these runs are shown in Fig.~\ref{fig:maxB}, left panel.

A simple scenario with a pure proton composition and an injection spectrum $\propto E^{-2}$ was simulated using CRPropa 3, for a maximal energy of $E_\text{max}=500 \; \text{EeV}$. Sources were assumed to follow the large-scale distribution of baryons in the cosmological simulations. Deflections were calculated for various energy bins, averaging over observers randomly placed within the simulation volume. This is shown in the right panel of Fig.~\ref{fig:maxB}.

As can be seen in Fig.~\ref{fig:maxB}, deflections of 60 EeV protons are $\simeq 5^\circ$ for runs F and S, $\simeq 10^\circ$ for L, and $\simeq 30^\circ$ for O for sources distant less than 10 Mpc from Earth. For the distance bin 10-20 Mpc, these numbers are $\simeq 7^\circ$, $14^\circ$, $7^\circ$, and $35^\circ$, respectively. These differences show that the power spectrum of cosmic magnetic fields is a source of uncertainty when studying arrival directions of cosmic rays. 

As a consequence of this study, even if voids are highly magnetized, deflections of protons from nearby sources are such that UHECR astronomy at energies $\gtrsim 60 \; \text{EeV}$ might be possible in a significant fraction of the sky, provided that the distribution of magnetic fields in the universe is not similar to that of run O, and that deflections in the galactic magnetic field are not much larger than in the JF 2012 model. 

\section{EGMFs from simulations with ENZO}

In Ref.~\cite{Hackstein:2016pwa} CRPropa has been used to simulate the propagation of UHE protons through EGMF models obtained with the MHD code ENZO~\cite{Bryan:2013hfa}. In addition to gravity and magneto-hydrodynamics, ENZO includes metallicity-dependent equilibrium gas cooling and star formation and feedback. It uses a particle-mesh N-body method to follow the dynamics of the Dark Matter and a variety of shock-capturing Riemann solvers to evolve the gas component. 

Both scenarios where the magnetization of the LSS is due to primordial magnetic fields and where it is due to astrophysical effects are tested in Ref.~\cite{Hackstein:2016pwa}. In this way the impact of different field strengths in clusters, filaments and voids are investigated. In all the obtained models the minimum magnetization level is much smaller than e.g. in Refs.~\cite{Sigl:2004yk,alvesbatista2017}. By choosing different Milky Way like halos as observer positions the effects of different local source distributions are investigated as well. 

The angular power spectra for different magnetic field models and different observer positions are shown in Fig.~9 of Ref.~\cite{Hackstein:2016pwa}. For $E>55$~EeV the predicted anisotropy is significantly more than the $99\%$ confidence level upper limit for isotropy for all multipole moments $l \leq 20$ for all EGMF models and most observer positions. However, the additional deflections by the GMF are not taken into account here and could make the UHECR sky more isotropic. A small subset of observers do give an UHECR distribution consistent with isotropy at the 2-3$\sigma$ level. This shows the importance of the local environment ($\ll 50$~Mpc) for determining the UHECR anisotropy. It can have a stronger influence on the observed cosmic-ray distribution than the tested magnetic field models.

Furthermore, these simulations show that a dipole component of $\gtrsim 10^{-2}$ coincides with a dominant source within 5~Mpc distance to the observer. This might be a hint for that a dominant nearby source (e.g. Centaurus A) is the origin of the observed dipole reported in Refs.~\cite{Aab:2016ban, ThePierreAuger:2014nja,Deligny:2015vol, Aab:2014ila}. 

Additionally, the energy dependence of the dipole and quadrupole moments in these simulations show a steepening of their slope at $\sim 40$~EeV. This coincides with the energy where the GZK threshold starts. Because cosmic rays with higher energies cannot reach us from farther away than $\sim 100$~Mpc only the local universe is seen at these energies. This increases the expected anisotropy in the UHECR sky.

By globally rescaling the magnetic field strength of one of the EGMF models the influence of the magnetization of voids is tested in Ref.~\cite{Hackstein:2016pwa} as well. They showed that in their simulations a deviation from isotropy occurs at energies of a few EeV for $l=2$ in primordial magnetic fields of $\leq 0.1$~nG. 

\section{UHECRs from Local Radio Galaxies}

The expectations from the observed local environment on UHECR arrival directions is investigated, using CRPropa, by Eichmann {\it et al.}~\cite{Eichmann:2017iyr}. They study whether the UHECRs we observe could come from local radio galaxies (RGs). As a source distribution the positions of the most luminous RGs within a distance of $\sim 120$~Mpc from the catalog of RGs from van Velzen {\it et al.}~\cite{vanVelzen:2012fn} have been implemented in CRPropa. The power of each source is scaled by the luminosity of the source. A distinction is made between two different source classes, namely RGs with jets and lobes and RGs without jets and lobes.

As EGMF model Eichmann {\it et al.} use the constrained model of Dolag {\it et al.}~\cite{Dolag:2004kp}. For Galactic deflections they use the lensing method of CRPropa with the JF 2012 model. Instead of only protons Eichmann {\it et al.} simulate $^1$H, $^4$He, $^{12}$C, $^{14}$N, $^{16}$O and $^{56}$Fe nuclei and fit their results to the mass composition and spectrum measurements by Auger. 

While their results reproduce the measured spectrum and chemical composition quite well, the observed low level of anisotropy is not obtained. This is because most UHECRs in such a local environment in this EGMF model do not suffer from significant deflections. Less than $1\%$ of all UHECR candidates show a deflection angle of $\theta_{\text{def}} \geq 90^{\circ}$. When the observed angular power spectrum is also included into the fit the anisotropy is reduced significantly, but still not enough to agree with the measured high degree of isotropy. Additionally, the observed chemical composition cannot be reproduced anymore.

\section{Conclusions}

With CRPropa 3~\cite{Batista:2016yrx} a tool has become available which allows for studying UHECR source scenarios including deflections in Galactic and extragalactic magnetic fields in an efficient way. In Ref.~\cite{alvesbatista2017} CRPropa has been used to investigate what can be expected in the case of maximum allowed intergalactic magnetic fields. This showed that, even if voids contain strong magnetic fields, deflections of protons with energies $\gtrsim 60 \; \text{EeV}$ from nearby sources might be small enough to allow for UHECR astronomy. 

In Ref.~\cite{Hackstein:2016pwa} several scenarios with a smaller magnetization have been studied using CRPropa. This showed that the local source distribution can have a more significant effect on the large-scale anisotropy than the EGMF model. For instance, a dipole component of $\gtrsim 10^{-2}$ as measured by Auger and TA could be explained by a dominant source within 5~Mpc distance. Furthermore, they showed that the average magnetic field in voids is most likely $\geq 0.1$~nG, otherwise a deviation from isotropy should have been observed at EeV energies.

To study if UHECRs could come from local RGs, Eichmann {\it et al.}~\cite{Eichmann:2017iyr} implemented a catalog of RGs as sources in CRPropa. They used the EGMF model of Dolag {\it et al.}~\cite{Dolag:2004kp}, the JF 2012 GMF model and fitted their results to the mass composition and spectrum measurements by Auger. In this setup they could not reproduce the observed low level of anisotropy. Therefore they concluded that the magnetic field strength in voids in the Dolag {\it et al.} EGMF is too low and/or there are additional sources of UHECRs that were not taken into account in these simulations.


\begin{thebibliography}{9}

\bibitem{Aab:2016ban} Pierre Auger Collaboration (A. Aab \textit{et al}.), {\it submitted to JCAP} (arXiv:1611.06812).
\bibitem{ThePierreAuger:2014nja} Pierre Auger Collaboration (A. Aab \textit{et al}.), Astrophys. J. \textbf{802}, 111 (2015).
\bibitem{Deligny:2015vol} O. Deligny for the Pierre Auger and Telescope Array Collaborations, PoS \textbf{ICRC2015}. 395 (2016).
\bibitem{Aab:2014ila} Pierre Auger and Telescope Array Collaborations (A. Aab \textit{et al}.), Astrophys. J. \textbf{794}, 172 (2014).
\bibitem{Kawata:2015whq} K. Kawata for the Telescope Array Collaboration, PoS \textbf{ICRC2015}, 276 (2016).
\bibitem{Abbasi:2014lda} Telescope Array Collaboration (R. U. Abbasi \textit{et al}.), Astrophys. J. \textbf{790}, L21 (2014).
\bibitem{Aublin:2015txt} J. Aublin for the Pierre Auger Collaboration, PoS \textbf{ICRC2015}, 310 (2016).
\bibitem{Aab:2016zth} Pierre Auger Collaboration (A. Aab \textit{et al}.), JCAP \textbf{1704}, 038 (2017).
\bibitem{Aab:2014kda} Pierre Auger Collaboration (A. Aab \textit{et al}.), Phys. Rev. \textbf{D90}, 122005 (2014).
\bibitem{Aab:2016vlz} Pierre Auger Collaboration (A. Aab \textit{et al}.), arXiv:1604.03637.
\bibitem{Batista:2016yrx} R. Alves Batista \textit{et al}., JCAP \textbf{1605}, 038 (2016).
\bibitem{Batista:2015mea} R. Alves Batista \textit{et al}., JCAP \textbf{1510}, 063 (2015).
\bibitem{Dolag:2004kp} K. Dolag \textit{et al}., JCAP \textbf{0501}, 009 (2005).
\bibitem{Prouza:2003yf} M. Prouza and R. Smida, Astron. Astrophys. \textbf{410}, 1 (2003). 
\bibitem{Sun:2007mx} X. H. Sun \textit{et al}., Astron. Astrophys. \textbf{477}, 573 (2008).
\bibitem{Pshirkov:2011um} M. S. Pshirkov \textit{et al}., Astrophys. J. \textbf{738}, 192 (2011).
\bibitem{Jansson:2012pc} R. Jansson and G. R. Farrar, Astrophys. J. \textbf{757}, 14 (2012).
\bibitem{Jansson:2012rt} R. Jansson and G. R. Farrar, Astrophys. J. \textbf{761}, L11 (2012).
\bibitem{Sigl:2004yk} G. Sigl, F. Miniati, and T. Ensslin, Phys. Rev. \textbf{D70}, 043007 (2004).
\bibitem{shin2017} M.-S. Shin, J. Devriendt and A. Slyz, {\it in preparation}, (2017).
\bibitem{alvesbatista2017} R. Alves Batista \textit{et al}., {\it submitted to PRD} (arXiv:1704.05869).
\bibitem{Hackstein:2016pwa} S. Hackstein \textit{et al}., Mon. Not. Roy. Astron. Soc. \textbf{462}, 3660 (2016). 
\bibitem{Bryan:2013hfa} G. L. Bryan \textit{et al}., Astrophys. J. Suppl. \textbf{211}, 19 (2014).
\bibitem{Eichmann:2017iyr} B. Eichmann \textit{et al}., arXiv:1701.06792.
\bibitem{vanVelzen:2012fn} S. van Velzen \textit{et al}., Astron. Astrophys. \textbf{544}, A18 (2012).

\end{thebibliography}
\end{document}